\documentclass[a4paper,11pt]{article}

\usepackage{amsmath}
\usepackage{amssymb}
\usepackage{color}
\usepackage[dvips]{graphicx}
\usepackage{cite}
\usepackage{hyperref}

\makeatletter
\@addtoreset{equation}{section}
\renewcommand{\theequation}{\thesection.\@arabic\c@equation}
\makeatother

\makeatletter
\renewcommand\appendix{\par%\newpage
  \setcounter{section}{0}%
  \setcounter{subsection}{0}%
  \gdef\thesection{Appendix \@Alph\c@section }
  \renewcommand{\theequation}
  {\Alph{section}.\arabic{equation}}
}
\makeatother

\def \be {\begin{equation}}
\def \ee {\end{equation}}
\def \ba {\begin{array}}
\def \ea {\end{array}}
\def \bea{\begin{eqnarray}}
\def \eea{\end{eqnarray}}

\def \d {\delta}

\def \m {\mu}
\def \n {\nu}
\def \k {\kappa}
\def \l {\lambda}

\def \r {\rho}

\def \f {\frac}

\def \nn {\nonumber}

\def \la {\leftarrow}

\def \sr {\sqrt}

\def \la {\label}
\def \fn {\footnote}

\title{\textbf{Emergent cosmic space in Rastall theory}}

\author{
Fang-Fang Yuan $^{1}$ \fn{kfyuan@yahoo.com} ,\,
Peng Huang $^{2}$ \fn{huangp46@mail.sysu.edu.cn}
}

\date{}

\begin{document}

\maketitle

\begin{center}
{\it
$^{1}$ Department of Physics, Hebei Normal University of Science and Technology, \\
 Qinhuangdao 066004, China  \\
\vspace{2mm}
$^{2}$ Department of Information, Zhejiang Chinese Medical University, \\
 Hangzhou 310053, China
}
\vspace{10mm}
\end{center}

\begin{abstract}

Padmanabhan's emergent cosmic space proposal is exploited to study the Rastall theory
which involves modifying the covariant conservation law of energy-momentum tensor.
As necessary elements for this approach,
we firstly find the Komar energy and the general entropy of apparent horizon in this theory.
After that, a modified expansion law is invoked to reobtain the Friedmann equations.

\end{abstract}

\baselineskip 18pt

\thispagestyle{empty}

\newpage

\section{Introduction}

Ever since the groundbreaking research of Jacobson \cite{Jacobson:1995ab},
the emergent gravity paradigm  has been investigated from various perspectives \cite{{Verlinde:2010hp},{Faulkner:2013ica}}.
%\cite{Padmanabhan:2014jta}.     % /spacetime
The original idea (from thermodynamics to gravity) is to use the Clausius relation and the properties of local Rindler horizon to derive the Einstein field equation.
%Furthermore, Verlinde  \cite{Verlinde:2010hp} has shown that the assumption of entropic force and the holographic principle
%lead to the second law of Newton and Newton's law of gravitation (see also \cite{Padmanabhan:2009vy}).
%One recent development of this approach (from thermodynamics to gravity) centers on the first law of entanglement entropy \cite{{Lashkari:2013koa},{Faulkner:2013ica},{Jacobson:2015hqa}}. % incarnation
On the other hand,
an inverse approach (from gravity to thermodynamics) has been proposed by Padmanabhan \cite{Padmanabhan:2002sha}
who found that
certain component of the gravitational field equations in a static, spherically symmetric spacetime
can be rewritten as an ordinary first law of thermodynamics at a black hole horizon.
Some aspects of this method are still under investigation \cite{{Kwon:2013dua},{Miao:2014wpa},{Chakraborty:2014joa},{Chakraborty:2015aja},{Chakraborty:2015hna}}.

As for the emergent cosmology (from thermodynamics to gravity),
Padmanabhan \cite{Padmanabhan:2012ik} has introduced an intriguing idea that
the expansion of the universe can be regarded as a process towards holographic equipartition.
This proposal has led to many applications and generalizations, see e.g. \cite{{Cai:2012ip},{Sheykhi:2013oac},{Yuan:2013nsa},{Chang-Young:2013gwa},{Sepehri:2014jla},{Sepehri:2016jfx}}.
In our work,
we plan to utilize this approach to study an interesting gravity called Rastall theory \cite{Rastall:1973nw}
which supposes that the energy-momentum is not covariantly conserved in curved spacetimes
\fn{ Over the years, Rastall's idea and formulation have stimulated many investigations
\cite{{Smalley:1974gn},{Fabris:2011wz},
{deMello:2014hra},{Carames:2014twa},{Salako:2016ihq}}.
Besides the applications in cosmology, Rastall theory can also be related to some modified gravity theories.
}.

As preparatory steps, we also find the Komar energy in this theory
and the general entropy of apparent horizon in the non-flat Friedmann-Robertson-Walker universe.
Obviously, these results are important in their own right.
As a side note, we would like to mention that
this paper is partially inspired by the works in \cite{{Moradpour:2016rml},{Moradpour:2016fur}}
which have studied the thermodynamics of apparent horizon in the flat case as well as
the relation between Rastall field equation and the first law.

The structure of this work is as follows.
In Section \ref{sec2},
we find the Komar energy/mass formula in Rastall theory.
In Section \ref{sec3},
we generalize the analysis of \cite{Moradpour:2016rml} to the case of non-flat universe,
and find the general entropy of apparent horizon.
By starting with a modified expansion law, in Section \ref{sec4},
we apply the emergent cosmic space proposal to Rastall theory and reproduce the Friedmann equations.
The conclusion will be given in the last section.

\section{Komar energy in Rastall theory}    \la{sec2}

Rastall \cite{Rastall:1973nw} proposed that one fundamental assumption of Einstein's general relativity,
i.e. the vanishing of covariant divergence of the energy-momentum, is questionable in curved spacetimes.
By supposing a relation as $  T^{\m\n}_{\ \ ;\n} = \l R^{,\m}  $
\fn{ We note in passing that Visser's massive graviton model \cite{Visser:1997hd} also involves a modification of the energy-momentum conservation equation.
},       % \nabla_\m
a generalization of the gravitational field equation is found to be
\be   \la{ras}
G_{\m\n} + k\l g_{\m\n} R = k T_{\m\n} .
\ee
Here $G_{\m\n}$ is the Einstein tensor and $k$ is the gravitational constant in Rastall theory
which is to determined by the consistency with Newtonian limit.
The Einstein field equation corresponds to the case where $\l = 0, k = 8\pi$.
Our convention for the Newton's constant is $G = 1$.

%----------------------------------------------------------------------

The Rastall field equation (\ref{ras}) can be rewritten as
\be
R_{\m\n} = k \bigg[ T_{\m\n} - \f{1-2k\l}{2(1-4k\l)} g_{\m\n} T \bigg] .
\ee
According to the standard procedure \cite{Komar:1958wp},
we see that the Komar energy/mass in Rastall theory has the following form
\fn{
One may try to add a prefactor $k/8\pi$ in this formula, but it would be inconsistent with our investigation in Section \ref{sec4}.
}:
\be
E_K =  \int_V dV \Big( 2T_{\m\n} - \f{1-2k\l}{1-4k\l} g_{\m\n} T \Big) u^\m \xi^\n .
\ee
Here $u^\m$ is a unit timelike vector and $\xi^\n$ is a normalized Killing vector.
Although this is an expected result, we are unaware of any previous detailed study on it in this particular theory.

Considering a perfect fluid again and choosing suitable coordinates, we can define the Komar energy density as (with $E_K = \r_K V$ for the usual case)
\bea    \la{Ked}
\r_K &=& 2T_{00} - \f{1-2k\l}{1-4k\l} g_{00} T  \nn  \\
&=&  \f{1}{1-4k\l} \big[ (1-6k\l) \r + 3(1-2k\l) p \big] .
\eea
Here $\r$ is the energy density and $p$ is the pressure.
This is the expression we will use in the following discussions.
One can also notice that the familiar equation $\r_K = \r+3p$ is obviously recovered when $\l = 0, k = 8\pi$.

%However, in some sense this formula has a conflict with the main characteristic of Rastall theory that the energy-momentum is not covariantly conserved.
%
%By inspection of the form of apparent horizon entropy (\ref{ent}),
%we argue that the physically relevant Komar energy density should be a modified one given by
%\be    \la{moK}
%\wtd m_K = m_K - \f{32\pi \l (1-3k\l)}{1-4k\l} \dot m_K .
%\ee
%This means that it can evolve with time and has a similar behavior to the entropy.
%Interestingly, this is also the unique expression to be consistent with our investigation in the next section.
%Of course, it would be desirable to have an independent and more rigorous derivation.

\section{Apparent horizon entropy: the non-flat case}       \la{sec3}

As shown in \cite{Moradpour:2016rml},
the method of Cai-Kim \cite{Cai:2005ra} can be invoked to obtain the apparent horizon entropy with spatial curvature factor $\k=0$.
The derivation relies heavily on Hayward's unified first law \cite{Hayward:1997jp}.

We would like to generalize this analysis to the non-flat Friedmann-Robertson-Walker universe.
In this case, we can use the Rastall field equation (\ref{ras}) to find the Friedmann equations for a perfect fluid as
\bea
3( 1-4k\l ) H^2 - 6k\l \dot H + 3 ( 1 - 2k\l ) \f{\k}{a^2} &=& k\r ,   \la{frie1}   \\
3( 1-4k\l ) H^2 + 2 ( 1 - 3k\l ) \dot H + ( 1 - 6k\l ) \f{\k}{a^2} &=& - k p  .  \la{frie2}
\eea
Here $H = {\dot a}/{a}$ is the Hubble parameter with $a$ the scale factor.
The location of the apparent horizon is as usual
\be     \la{loc}
r_A = \f{1}{\sr { H^2 + \f{\k}{a^2} } } .
\ee

The energy flux crossing the apparent horizon assumes a familiar form:
\be
\d Q = - 3 V H (\r+p) dt .
\ee
Using the Clausius relation, the variation of entropy is given by
\be     \la{vaent}
d S = - \f{\d Q}{T} % = 6\pi V (\r+p) H r_A dt
= 8 \pi^2 H r^4_A (\r+p) dt .
\ee
Recalling the following continuity equation ($\m = 0$ component of the nonconservation relation $T^{\m\n}_{\ \ ;\n} = \l R^{,\m}$)
\be   \la{con}                      % which is just the
\f{1-3k\l}{1-4k\l} \dot \r - \f{3k\l}{1-4k\l} \dot p + 3H (\r+p) = 0 ,
\ee
then (\ref{vaent}) can be rewritten as
%\be
%d S = - \f{8 \pi^2}{3 (1-4k\l)} r^4_A \big[ (1-3k\l) d\r - 3k\l dp \big] .
%\ee
\be
\f{dS}{dt} = - \f{8 \pi^2}{3 (1-4k\l)} r^4_A \big[ (1-3k\l) \dot \r - 3k\l \dot p \big] .
\ee

Now we insert the Friedmann equations (\ref{frie1}) and (\ref{frie2}) into the above expression, and obtain a simple result
\be
\f{dS}{dt} = - \f{16\pi^2}{k} H r^4_A \Big( \dot H - \f{\k}{a^2} \Big) .
\ee
Noticing that $\dot r_A = - H r^3_A \Big( \dot H - \f{\k}{a^2} \Big)$,
then the apparent horizon entropy is obtained as follows
\be    \la{ent}
S = \f{16\pi^2}{k} \int r_A d r_A = \f{8\pi^2}{k} r^2_A .
\ee
This obviously has the familiar form: $S = {2\pi A}/{k}$,
where $A = 4\pi r^2_A$ is the area of apparent horizon
\fn{
Note that the derivation in \cite{Moradpour:2016rml} has an unfortunate mistake (compare (\ref{ent}) here with (22) in that reference).
This can be most easily seen by inserting the Raychaudhuri equation (9) there into (17) of \cite{Moradpour:2016rml}.
In an earlier version of our present work on arXiv, we argued for a modified Komar energy % to be consistent with the entropy (22) of \cite{Moradpour:2016rml}.
which now turns out to be unnecessary because of this error.
Accordingly, a lot of complicated and tedious calculations including an unsuccessful attempt on the non-flat case would have been avoided for an acute reader.
}.
Surprisingly, the main result (\ref{ent}) of this section is independent of the parameter $\l$,
and has no entropic correction in Rastall theory.

%\bea
%\f{d\bar V}{dt} &=& \f{r_A}{2} \f{d\bar A}{dt} = 2 r_A \f{d\bar S}{dt}    \nn  \\
%&=& 2 H r^2_A \f{dS}{dt} + \f{24\pi \k }{k} \f{r^2_A \dot a}{a^3}      \nn   \\
%&=& H^2 r^2_A \f{d\wtd V}{dt} - \f{12\pi}{k} r^2_A \f{d}{dt} \Big( \f{\k}{a^2} \Big)
%\eea
%\be
%\bar N_{sur} = H^2 r^2_A \wtd N_{sur} +
%\ee

%\be
%m_K = - \f{3}{4 \pi} \f{H^2}{r^2_A} ( H^2+\dot H )
%\ee

%\be
%m_K = - \f{3}{4 \pi} \bigg( H^2+\dot H + \f{1-2k\l}{1-4k\l} \f{\k}{a^2}  \bigg)
%\ee

\section{Friedmann equations from an expansion law}     \la{sec4}

According to the emergent cosmic space proposal of Padmanabhan \cite{Padmanabhan:2012ik},
the expansion of the universe can be attributed to the difference between the number of degrees of freedom on a holographic
surface $N_{sur}$ and the one in the bulk $N_{bulk}$.
Thus its expansion is interpreted as a process towards holographic equipartition.
In the case of non-flat universe, a modified expansion law has been proposed in \cite{Sheykhi:2013oac} as
\be     \la{mlaw}
\f{d V}{dt} = H r_A (  N_{sur} - N_{bulk} ) .
\ee
This is also applicable to the situations with entropic corrections \cite{Yuan:2013nsa}.
%Besides $N_{bulk}$, other two parameters have been modified as compared to the original formula.
%This is because the apparent horizon entropy deviates from the standard holographic situation where one has an area law as $S=A/4$.
%Therefore, to agree with the holographic principle, we need to introduce an effective area to preserve the area law.

%\be
%dS = - \f{16\pi^2}{k} \f{dH}{H^3} + \f{32\pi^2 \l (1-3k\l)}{1-4k\l} \f{d\dot H}{H^4}
%\ee
Using the expression of apparent horizon entropy in (\ref{ent}), we can obtain the increase in the effective cosmic volume as
\bea     \la{effv}
\f{d V}{dt} &=& \f{r_A}{2} \f{d A}{dt} = 2 r_A \f{dS}{dt}    \nn  \\
&=& - \f{32\pi^2}{k} r^5_A H \Big( \dot H - \f{\k}{a^2} \Big)    \nn    \\
&=& - 2\pi r^5_A \f{d}{dt} \bigg[ \f{8\pi}{k} \Big( \f{1}{r^2_A} \Big)  \bigg] .
\eea
This result and the holographic idea invite us to propose the following effective number of degrees of freedom on the apparent horizon
\be     \la{effs}
N_{sur} = 4\pi r^2_A \cdot \f{8\pi}{k} .
\ee
When $\k = 0, k = 8\pi$,  we have $N_{sur} = {4\pi}/{H^2}$ which corresponds to the standard holographic situation \cite{Padmanabhan:2012ik}
(with the Planck length $L_P = 1$).
In this special case, the effective volume reduces to the usual Hubble volume.

Recalling the Komar energy density in (\ref{Ked}),
we find a general formula for the modified number of degrees of freedom in the bulk as
\bea      \la{effb}
N_{bulk} &=& \f{2}{T} | E_K | = - 2 \r_K \f{V}{T}     \nn  \\
&=& - \f{16 \pi^2}{3} r^4_A \cdot \r_K .
\eea
Here the cosmic volume is $V = 4\pi r_A^3 / 3$, and the Hawking temperature is $T = {1} / {2\pi r_A}$ \cite{Cai:2008gw}.
Moreover, the appearance of a minus sign is because we are considering the accelerating phase.

%-----------------------------------------------------------------

Based on the results in (\ref{effv}) and (\ref{effs}),
we can find that the modified expansion law (\ref{mlaw}) leads to
\be
\r_K = - \f{3}{16 \pi^2 r^4_A} \bigg[  4\pi r^2_A \cdot \f{8\pi}{k} + \f{32\pi^2}{k} r^4_A \Big( \dot H - \f{\k}{a^2} \Big) \bigg] .
\ee
Together with the radius of apparent horizon (\ref{loc}),
we then have an elegant result
\be     \la{mKH}
\r_K = - \f{6}{k} (H^2+\dot H) .
\ee

Recalling the detailed formula of the original Komar energy density in (\ref{Ked}),
the above formula enables us to express the pressure as
\be
p = \f{1}{3(1-2k\l)} \Big[ - \f{6}{k} (1-4k\l) (H^2+\dot H) - (1-6k\l) \r \Big] .
\ee
We then insert this into the continuity equation (\ref{con}).
After some manipulations and eliminating a common factor ${1}/{(1-2k\l)}$, we obtain
\be        \la{main}
\dot \r + 2 H \r - \f{6}{k} (1-4k\l) H (H^2+\dot H) + 6 \l (2H\dot H+\ddot H) = 0 .
\ee
Although the appeal to continuity equation may raise some suspicion,
it is actually a routine step in this research area.
Obviously (\ref{con}) can reduce to a familiar equation as $\dot \r + 3H (\r+p) = 0 $.

Multiplying the above equation by $a^2$, we obtain
\be
\f{k}{3(1-4k\l)} ( a^2 + 2 a \dot a \r ) = 2 a^2 H \Big( H^2 - \f{2k\l}{1-4k\l} \dot H \Big) + a^2 \Big( 2H\dot H - \f{2k\l}{1-4k\l} \ddot H \Big) .
\ee
This can be rewritten as
\be
\f{k}{3(1-4k\l)} \f{d}{dt} ( \r a^2 ) = \f{d}{dt} \Big[ a^2 \Big( H^2 - \f{2k\l}{1-4k\l} \dot H \Big) + C \Big] .
\ee
Here we have an integration constant $C$ which is not necessarily zero.
Through the comparison with the Friedmann equation (\ref{frie1}),
one can inspect that by adjusting the arbitrary
constant $C$,
these two equations agree even when the curvature factor $\k\neq 0$.
%\fn{
%Inspired by the experience in \cite{Yuan:2013nsa}, one may try to avoid this problem by replacing the derivatives of $H$ by $r_A$
%in the Friedmann equations.
%This involves the following expressions:
%\be
%H^2 = \f{1}{r^2_A} - \f{\k}{a^2} ,   \quad \dot H = - \f{\dot r_A}{H r^3_A} + \f{\k}{a^2} ,
%\ee
%%\be
%%\dot H = - \f{\dot r_A}{H r^3_A} + \f{\k}{a^2} ,
%%\ee
%\be
%\ddot H = - \f{\ddot r_A}{Hr^3_A} - \f{\dot r_A^2}{Hr^4_A} \Big( \f{1}{H^2r^2_A} - 3 \Big) +
%\f{\dot r_A}{H^2r^3_A} \f{\k}{a^2} - 2 H \f{\k}{a^2} .
%\ee
%However, this procedure doesn't do the trick either.
%}.
We note in passing that to derive this result, one may also use the invariant volume instead of the above aerial volume as in \cite{Chang-Young:2013gwa}.

%The reason we didn't get the original Friedmann equation (\ref{frie1}) may be attributed to the following fact:
%\be
%\f{d}{dt} [ a^2 \cdot 3(1-2k\l) \f{\k}{a^2} ] = \f{d}{dt} const. \equiv 0 .
%\ee
%Recalling the formula in (\ref{mKH}), one can also obtain the other Friedmann equation as follows
%\be
%3( 1-4k\l ) H^2 - 2 ( 1 - 3k\l ) \dot H = - k p .
%\ee
%In hindsight, if the Friedmann equations were given at first,
%then the "derivation" of equations like (\ref{mKH}) and (\ref{main}) would also
%involve some interesting cancellations of the complicated coefficients.
%It is important to note that even though we didn't get the Friedmann equations with $\k\neq 0$,
%we still have to use the modified expansion law (\ref{mlaw}) rather than the original formula.
%Otherwise, we would not get the key formula in (\ref{mKH}).

To sum up, in the context of an unordinary gravity like Rastall theory,
the basic expansion law can still lead to the fundamental equations for the expansion of universe.

\section{Conclusion}

In this work, we firstly propose a Komar energy density (\ref{Ked}) for a perfect fluid in Rastall theory.
After that, we generalize the study in \cite{Moradpour:2016rml} to the non-flat universe,
and find the entropy of apparent horizon in (\ref{ent}).
Based on these analyses,
we then apply a modified expansion law (\ref{mlaw}) in the emergent cosmic space proposal to Rastall theory,
and recover the basic Friedmann equation in (\ref{frie1}).
Because of an arbitrary integration constant, the derivation here is independent of the value of curvature factor.
Notice also that
the consistency for the case with $r_A \neq 1/H$ requires us to use a modified expansion law like (\ref{mlaw}).

As for further investigations,
it would be valuable to extend this work to the Brans-Dicke-Rastall theory \cite{Carames:2014twa}.
Secondly, one may consider the case of event horizon in Rastall theory along the line of \cite{{Saha:2012nz},{Mitra:2015jqa}}.
Thirdly, \cite{Moradpour:2016fur} has recently shown the relation between Rastall field equation and the ordinary first law of thermodynamics.
By using the method of \cite{{Kwon:2013dua},{Miao:2014wpa}}, maybe the exact first law of black hole mechanics (with black hole charges) could also be obtained in this theory.
Finally,
the idea that the energy-momentum is not covariantly conserved in cosmology
perhaps could be combined with the recent proposal of uniformly expanding vacuum \cite{Huang:2016fws}.

%\Acknowledgments
%\noindent
{\large{\bf Acknowledgments}}

We are indebted to an anonymous referee for valuable and meticulous suggestions.
FFY is partially supported by the Doctoral Foundation of HNUST; PH is supported by the Doctoral Foundation of ZCMU.

%\vspace*{5mm}

\end{document}